\documentstyle[aps]{revtex}

\begin{document}

\title{Two-dimensional quantum interference contributions
to the magnetoresistance of Nd$_{2-x}$Ce$_{x}$CuO$_{4-\delta}$ single crystals}

\author{G.I.~Harus, A.N.~Ignatenkov, A.I.~Ponomarev, 
L.D.~Sabirzyanova,\\ N.G.~Shelushinina}
\address{Institute of Metal Physics\\ 18 Kovalevskaya St., Ekaterinburg, 620219,
Russia}

\author{N.A.~Babushkina}
\address{RSC Kurchatov Institute 123182 Moscow, Russia}

\maketitle

\begin{abstract}
The $2D$ weak localization effects at low temperatures $T = (0.2\div 4.2)\,K$ 
have been investigated in nonsuperconducting sample 
Nd$_{1.88}$Ce$_{0.12}$CuO$_{4-\delta}$ and in the normal state of the 
superconducting sample Nd$_{1.82}$Ce$_{0.18}$CuO$_{4-\delta}$ for 
$B~>~B_{c2}\simeq 3\,T$. The phase coherence time $\tau_{\varphi}(\simeq 5 \cdot 
10^{-11}\,s$ at $1.9\,K$) and the effective thickness of a conducting $CuO_2$ 
layer $d (\simeq 1.5\,$\AA)\ have been estimated by the fitting of $2D$ weak 
localization theory expressions to the magnetoresistivity data for the normal 
to plane and the in-plane magnetic fields. The estimation of the parameter $d$ 
ensures the condition of a strong carrier confinement and makes a basis to the 
model of almost decoupled $2D$ metallic sheets for the 
Nd$_{2-x}$Ce$_{x}$CuO$_{4-\delta}$ single crystals.

\end{abstract}

\section{Introduction}
The crystallographic structure $T'$ of Nd$_{2-x}$Ce$_{x}$CuO$_{4-\delta}$ is 
the simplest among the superconducting cuprates, each copper atom is 
coordinated to four oxygen atoms in a planar structure without apical oxygen. 
A Nd$_2$CuO$_4$ crystal is the insulator with the valence band to be mainly of 
$O\ 2p$ character and the empty conduction band to be the upper Hubbard 
$Cu\ 3d$ band. The Coulomb $3d-3d$ repulsion on $Cu$ site 
$U\ (\simeq 6\div 7eV)$ is strong and it is larger than oxygen to metal 
charge-transfer energy $D\ (\simeq 1\div 2 eV)$. Thus these cuprates are 
classified as the charge-transfer semiconductors.

The combination of $Ce$ doping and $O$ reduction results in the $n$-type 
conduction in the $CuO_2$ layers. The energy band structure calculation [1] 
shows that the Fermi level is located in the band of $pd\sigma$-type formed by 
$3d(x^2-y^2)$ orbitals of $Cu$ and $p_{\sigma}(x,y)$ orbitals of oxygen. 
The $pd\sigma$\ band appears to be of highly two-dimensional character with 
almost no dispersion in the normal to $CuO_2$ planes $z$-direction. The 
electrons are concentrated within the confines of conducting $CuO_2$ layers 
separated from each other by a distance $c\simeq 6\,$\AA.
 
Due to the layered crystal structure the high-T$_c$ copper oxide compounds are 
highly anisotropic in their normal state electical properties. The 
electron-doped systems Nd$_{2-x}$Ce$_{x}$CuO$_{4-\delta}$ exhibit a very 
large anisotropy factor, $\rho_c/\rho_{ab}\geq 10^4$ [2,3] that is somewhat 
lower than in Bi-systems ($\rho_c/\rho_{ab}\sim 10^5$) but essentially higher 
than in La- and Y-systems ($\rho_c/\rho_{ab}\sim 10^2$). For the underdoped and 
optimally doped compounds the $c$-axis resistivity, $\rho_c$, is usually 
non-metallic ($d\rho_c/dT\le\,0$) at low enough temperatures [4].
In contrast to it, the magnitude and the temperature dependence of the 
resistivity in a $CuO_2$ plane, $\rho_{ab}$, are in general metallic near 
optimum doping.

A two-dimensional metallic state in a system with random disorder should
exhibit weak localization of the charge carriers at low temperatures [5].
Weak localization behaviour of the in-plane resistivity has been clearly
observed and perfectly analysed for the Bi$_2$Sr$_2$CuO$_6$ systems which were 
investigated with high precision at $T$ down to $0.5\,K$ in normal and 
perpendicular to the $CuO_2$ planes magnetic fields up to $8\,T$ [6]. As for 
La$_{2-x}$Sr$_{x}$CuO$_{4-\delta}$ or La$_{2-x}$Ba$_{x}$CuO$_{4-\delta}$ 
systems, a concentration range between the hopping regime at low $x$ and 
superconducting regime at $x>0{.}05$ seems
to be so narrow that a well defined weak localization behaviour is difficult
to observe [7,8]. Only in the close proximity to $x = 0{.}05$ nonsuperconducting
sample La$_{2-x}$Ba$_{x}$CuO$_{4-\delta}$ [7] and superconducting sample 
La$_{2-x}$Sr$_{x}$CuO$_{4-\delta}$ ($T_c = 4\,K$) in the fields $B > 8\,T$ 
display some signs of weak localization ($ln T$-dependence of $\rho_{ab}$).

Due to their $T'$ structure the $Nd$-systems should be particularly 
advantageous for the observation of $2D$ effects in conduction process. 
Really, there are several reports on the manifestation of $2D$ weak 
localization in the in-plane conductance of 
Nd$_{2-x}$Ce$_{x}$CuO$_{4-\delta}$ single crystals or films. Thus a linear 
dependence of resistivity on $lnT$ comes about at $T < T_c$ for samples with 
$x\cong 0{.}15$, in which superconducting state is destroyed by a magnetic 
field [9]. Furthermore, a highly anisotropic (with regard to the magnetic 
field direction) negative magnetoresistance, predicted for $2D$ weak 
localization, has been observed in the nonsuperconducting state at low 
temperatures: in highly underdoped sample with $x = 0{.}01$ [10] and in 
unreduced samples with $x = 0{.}15$ [11] or $x = 0{.}18$ [12]. Measurements 
in superconducting $x=0{.}15$ sample with high $T_c$ ($T_c = 20\,K$) has shown 
a similar negative magnetoresistance in high (up to $30\,T$) transverse 
magnetic fields and an upturn in the normal state resistance as $T$ is 
lowered [11].

In our previous investigation of the sample with $x=0{.}18$ ($T_c=6\,K$) a 
negative magnetoresistance has been observed after the destruction of 
superconductivity by a magnetic field up to $5{.}5\,T$ at $T\leq\,1{.}4\,K$ 
[13]. We report here the results of measurements at much lower temperatures 
(down to $0{.}2\,K$) and in the higher {\it dc} magnetic fields (up to $12\,T$). 
A drastic dependence of magnetoresistance magnitude on the direction of 
magnetic field is the most important experimental test for the $2D$-character 
of a conducting system. For the investigation of the magnetoresistance 
anisotropy we have used here the measurements on nonsuperconducting sample with 
$x$ value ($x=0{.}12$) which is close to the boundary value $x=0{.}14$ for the 
superconductivity in a Nd$_{2-x}$Ce$_{x}$CuO$_{4-\delta}$ system.

\section{Results} 
High-quality single-phase Nd$_{2-x}$Ce$_{x}$CuO$_{4-\delta}$ 
($x=0{.}12\div\,0{.}20$) thin films have been produced by modificated 
lazer deposition technique with a flux separation [14]. The films with 
thickness around 5000\AA\ were deposited onto hot single crystal SrTiO$_2$ 
substrate, which has (100) surface orientation. It was necessary to anneal the 
films subsequently in vacuum $<\,10^{-2}$ torr at 800$^\circ\,C$ during 40 min 
to form superconducting phase. The X-ray diffraction study has revealed the 
existence of the tetragonal phase only with $c$-axis perpendicular to the film 
plane. We report here the data for Nd$_{2-x}$Ce$_{x}$CuO$_{4-\delta}$ films 
with $x=0{.1}2$ and 0.18 only.

The in-plane resistivity $\rho_{ab}$ and Hall coefficient $R$ 
($\vec{j}\|ab, \vec{B}\|c$) have been investigated in a single crystal 
superconducting film Nd$_{1.82}$Ce$_{0.18}$CuO$_{4-\delta}$ ($T_c = 6\,K$) 
at $T = (0.2\div 20)\,K$ in a magnetic field up to $B = 12\,T$. In the 
superconducting sample the normal state transport at low $T$ is hidden unless 
a magnetic field $B$ higher than the second critical field $B_{c2}$ is applied 
(for $B\perp ab\quad B_{c2}\cong 3\,T$ at $T = 4.2\,K$). We have destroyed 
superconductivity by a magnetic field perpendicular to $CuO_2$ planes and 
observed a negative magnetoresistance in fields higher than $B_{c2}$ (Fig.1) 
with logarithmic temperature dependence of the resistivity at $T < 4.5\,K$ 
(Fig.2). Fig.3 shows the results of measurements of the in-plane conductivity 
in non-superconducting sample Nd$_{1.88}$Ce$_{0.12}$CuO$_{4-\delta}$ for 
perpendicular $B_{\perp}$ and parallel $B_{\|}$ to the $CuO_2$ planes magnetic 
fields up to $B = 5.5\,T$ at $T=1.9\,K$ and $4.2\,K$.

\section{Discussion}
The logarithmic low temperature dependence of the conductivity is one of the 
indications of the interference quantum correction due to weak localization 
or electron-electron interaction in a two-dimensional system. Magnetic field 
normal to the motion of a carrier destroys the interference leading to the 
localization. In $2D$ system it causes negative magnetoresistance for the 
field perpendicular to the plane but no effect for the parallel configuration. 
In the $2D$ weak localization theory the quantum correction to the Drude 
surface conductivity in a perpendicular magnetic field is given by [15]:
\begin{equation}
\Delta\sigma_s(B_{\perp})=\alpha\frac{e^2}{\pi h}\cdot
\biggl\{\Psi\Bigl(\frac{1}{2}+\frac{B_{\varphi}}{B_{\perp}}\Bigr)-
\Psi\Bigl(\frac{1}{2}+\frac{B_{tr}}{B_{\perp}}\Bigr)\biggr\}  \label{eq:s_1}
\end{equation}
where $\alpha$ is a prefactor of the order of unity, $\Psi$ is the digamma 
function, $B_{\varphi}=c\hbar/4eL^2_{\varphi}$ and $B_{tr}=c\hbar/2e\ell^2$.
Here $L_{\varphi}=\sqrt{D\tau_{\varphi}}$ is the phase coherence length 
($D$ is the diffusion coefficient and $\tau_{\varphi}$ is the phase breaking
time) and $\ell$ is the mean free path. At low temperature the inequality 
$B_{\varphi}\ll\,B_{tr}$ ($L_{\varphi}\gg\,\ell$) is valid and thus the weak 
localization effects is almost totally supressed for $B\cong B_{tr}$. Let us 
compare the equation for the transport field, presented in the form
\begin{equation}
2\pi\cdot B_{tr}\ell^2=\Phi_0   \label{eq:B_tr}
\end{equation}
where $\Phi_0 = \pi c\hbar/e$ is the elementary flux quantum, with the 
relation between the coherence length $\xi$ and the second critical field 
in the so called ``dirty'' limit ($\xi\gg\ell$):
\begin{equation}
2\pi\cdot B_{c2}\ell\xi=\Phi_0.     \label{eq:B_c2}
\end{equation}
From Eqs~(2) and (3) one has $B_{tr}/B_{c2} = \xi/\ell$ and thus the inequality
$B_{tr}\gg B_{c2}$ should be valid for any dirty superconductor.

\subsection{Superconducting sample ($x = 0.18$)}
From the experimental values of $\rho_{ab}$ and Hall constant $R$ in the normal 
state we have obtained the Drude conductivity of a $CuO_2$ layer 
$\sigma_s = (\rho_{ab}/c)^{-1}$, the bulk $n = (eR)^{-1}$ and the surface 
$n_s = nc$\ \ electron densities ($c = 6\,$\AA\ is the distance between $CuO_2$ 
layers). We have $\sigma_s = 10^{-3}\Omega^{-1}$, $n = 1.1\cdot 10^{22} cm^{-3}$ 
and $n_s = 6.6\cdot 10^{14} cm^{-2}$ at $T=4.2\,K$ and $B>B_{c2}$. Using the 
relation $\sigma_s = (e^2/h)k_F\ell$, with $k_F$ to be the Fermi wave vector, 
we have estimated the parameter $k_F\ell\cong 25$. As $k_F\ell\gg 1$ a true 
metallic conduction in $CuO_2$ layers takes place. 

Since $k_F = (2\pi n_s)^{1/2}\cong 6\cdot 10^7 cm^{-1}$ we have found the mean 
free path $\ell\cong 4\cdot 10^{-7} cm$ and according to Eq. (2) the transport 
field $B_{tr}\simeq 20\,T$. In the investigated sample the second critical 
field $B_{c2}\simeq 3\,T$ at $T = 4.2\,K$\ [13] and $B_{c2}\simeq 5\,T$ at 
$T=0.2\,K$. Thus we have $B_{tr}\gg B_{c2}$ and it occurs possible to observe 
the negative magnetoresistance owing to $2D$ weak localization in the interval 
of magnetic fields $B_{c2} < B_{\perp} < B_{tr}$\ (Fig.1a).

In the field range $B_{\varphi}\ll B\ll B_{tr}$ the expression (1) may be 
writen as
\begin{equation}
\Delta\sigma_s(B_{\perp})=\alpha\frac{e^2}{\pi h}\cdot
\biggl\{-\Psi\Bigl(\frac{1}{2}\Bigr)-
ln\frac{B_{\perp}}{B_{tr}}\biggr\}.  \label{eq:s_2}
\end{equation}
Fig.1b shows the surface conductivity $\sigma_s$ as a function of $lnB$. It 
is seen that at $B > B_{c2}$ the experimental data can be described by simple 
formula (4) with prefactor $\alpha$ as the only fitting parameter. At 
$T=2\,K$ we have $\alpha = 1.5$ but as the temperature is lowered the 
$\alpha$~value becomes essentially more than unity: 
$\alpha = 6{.}6$ at $T = 0{.}2\,K$. Thus the effect of negative 
magnetoresistance at the lowest temperature is too large to be caused by the 
supression of weak localization only.

In the case of effective electron attraction there exists the other orbital
contribution to the negative magnetoresistance, namely, the contribution due
to disorder-modified electron-electron interaction in the so called Cooper
channel (the interaction of electrons with the opposite momenta) [16]. The 
contribution such as that may be the reason of the extra effect of negative 
magnetoresistance in our {\it in situ} superconducting sample at very low 
temperatures. The magnitudes of the coefficient of $lnB$ in superconducting 
aluminium films at $T > T_c$ have been quantitatively explained just so [17].

When the magnetic field is applied parallel to the $ab$-plane, it turned out 
that the upper critical field, $B^{\|}_{c2}$, is too high and is not 
reached in our sample with $x = 0.18$ in a fields up to $B = 12T$. This result 
is in accordance with the observation of a large anisotropy of $B_{c2}$\ for 
Nd$_{2-x}$Ce$_{x}$CuO$_{4-\delta}$ single crystals with $x = 0.16$ 
($B^{\perp}_{c2}= 6.7\,T,\ B^{\|}_{c2}=137\,T$) [18]. Thus, in order to 
investigate the dependence of magnetoresistance on the direction of magnetic 
field relative to $CuO_2$\ plane, the study of {\it in situ} nonsuperconducting 
sample is needed.

\subsection{Nonsuperconducting sample ($x = 0.12$)}
The positive magnetoconductivity (negative magnetoresistance) observed for 
this sample is obviously anisotropic relative to the direction of magnetic 
field (see Fig.~3). From the fit of the curves $\sigma_s(B_{\perp})$ by the 
functional form (1) (solid curves in Fig.~3) we have found the inelastic 
scattering length $L_{\varphi}=550$\AA\quad at $\ T=1.9\,K$ and 
$L_{\varphi}=770$\AA\ at $T = 4.2\,K$. For the in-plane diffusion coefficient 
$D_{\|}=(\pi\hbar^2/me^2)\sigma_s$ we have $D_{\|}=1.1 cm^2/s$, so that 
$\tau_{\varphi} = 5.4\cdot 10^{-11}s$\quad at $\ T=1.9\,K$ and 
$\tau_{\varphi} = 2.7\cdot 10^{-11}s$\quad at $\ T=4.2\,K$. These values are of the 
same order of magnitude as that obtained by Hagen et al. for $x\simeq  0.01$ 
crystal ($\tau_{\varphi} = 1.2\cdot 10^{-11}s$\ at $T = 1.6\,K$) [10], but in 
contrast to their unusual $\tau_{\varphi}\sim T^{0.4}$ dependence at 
$T< 10\,K$ our data at $T = 1.9\,K$ and $T=4.2\,K$ are compatible with the 
$\tau_{\varphi}\sim T^{-1}$ dependence, predicted for the electron-electron 
inelastic scattering in a disordered $2D$ system [19].

Much more weak negative magnetoresistance for parallel configuration 
($B\| ab$) is quadratic in $B$ up to $B_{\|} = 5.5\,T$ (see solid curves in 
Fig.~3). It is of the same order of magnitude as that of Hagen et al. at 
$T< 5\,K$ [10] or Kussmaul et al. at $T\leq 4.2\,K$ [11] but we havn't seen 
any sign of a positive kink at $B=(1\div 1.5)\,T$ observed in [10].

Longitudinal magnetoresistance in a strictly $2D$ system may be 
caused only by the influence of the field on the spin degrees of freedom. 
One of the most obvious reason for negative magnetoresistance is the 
scattering of electrons on some spin system: the system of $Cu$ spins or 
partially polarized $Nd$ spins. For any source of spin scattering the field 
scale for $B^2$ dependence ($B\ll B_s, B_s = kT/g\mu_B$) is too low to explain 
our experimental data. For $g\,=\,2\quad B_s\,=\,1{.}5\,T$ at $T\,=\,1.9\,K$ 
and $B_s = 3\,T$ at $T = 4.2\,K$, but we observe no deviations from $B^2$ 
dependence up to $B = 5.5\,T$. 

In standard theory of quantum interference effects in disordered conductors 
[20,21] the isotropic contribution to magnetoconductivity associated with 
spin degrees of freedom also takes place. When the Zeeman energy of electrons 
$g_e\mu_B B$ exceeds $kT$, the magnetic field suppresses the contribution to
conductivity originated from the part of electron-electron interaction thus 
leading to the effect of magnetoresistance. But this magnetoresistance is 
always positive and, with the value $g_e = 2$ for the electronic $g$-factor, 
has the same characteristic field as that for spin scattering: $B = B_s$. 
Thus it apparenly is not related to the effect in question, but it may be 
a reason of positive kink on magnetoresistance curves of Hagen et al. [10].

It is very important that in quasi-two-dimensional system with finite 
thickness $d~\ll~L_{\varphi}$ there exists an orbital contribution to the 
longitudinal magnetoresistance. It is an ordinary explanation for the 
parabolic negative magnetoresistance observed in parallel configuration in 
semiconducting $2D$ system: in $GaAs/AlGaAs$ heterostructures [22] or in 
silicon inversion layers [23]. The finite thickness correction to the strictly 
$2D$ theory is defined by the expression  [24]:
\begin{equation}
\Delta\sigma_s(B_{\|})=\frac{e^2}{\pi h}\ 
ln{\biggl[1+\biggl(\frac{B}{B^{\ast}}\biggl)^2\biggr]}, \label{eq:s_P}
\end{equation}
where $B^{\ast}=\sqrt{3}c\hbar/edL_{\varphi}$. It is seen from Eq.(5) that 
$\sigma_s(B_{\|})$ should be quadratic in $B$ at fields $B\ll B^{\ast}$ with 
characteristic field $B^{\ast}\cong (L_{\varphi}/d)B_{\varphi}\gg B_{\varphi}$.

For a preliminary estimation of $B^{\ast}$ let us assume that $d < c$ 
($c = 6$\AA\ is the distance between adjacent $CuO_2$ planes), then we have 
$B^{\ast} > 25\,T$\ at\ $T = 1.9\,K$ and $B^{\ast} > 35\,T$\ at\ $T = 4.2\,K$. 
Thus we think (so as Kussmaul et al. [11]) that the finite thickness correction 
to the $2D$ weak localization effect can reasonably explain the observed 
negative magnetoresistance for parallel configuration. As well as in parallel 
configuration there exists the finite thickness correction to the basic effect 
in perpendicular magnetic field (see [20], p.109). We believe that it is just 
the reason of the upturn of experimental points for $\sigma_s(B_{\perp})$\ at 
$B > (3.5\div 4)\,T$ in Fig.3. By fitting of the 
theoretical expression (5) to the curves for $\sigma_s(B_{\|})$ and taking 
into account the values of $L_{\varphi}$ obtained earlier, we have found for 
the effective thickness of a conducting $CuO_2$ layer $d = (1.5\pm 0.5)$\AA. 

The value of $d$ gives an estimate for the dimension of electron wave 
function in the normal to a $CuO_2$ plane direction and ensures the condition 
of a strong carrier confinement: $d < c$. It is in accordance with the proposed 
highly $2D$ character of the actual electron band of $pd\sigma$-type with 
almost no dispersion along $c$-axis [1]. The X-ray investigations also show 
the concentration of electron density within the limits of $\pm 1$\AA\ above 
and below a $Cu$ atom in $c$-direction [25]. The single crystal NdCeCuO may 
therefore be regarded as multi-quantum-well system ($1.5$\AA\ wells / $4.5$\AA\ 
barriers) or as an analog of multi-layered heterostructure. The theoretical 
description of high-$T_c$ superconductors as heterostructures has been 
recently proposed [26].

As the $2D$-version of weak localization theory is able to describe the 
behaviour of $\sigma_s(B,T)$ in our sample, the inequality 
$\tau_{esc} > \tau_{\varphi}$ should be valid for the escape time of electron 
from one $CuO_2$ plane to another. The escape time between adjacent quantum 
wells in multilayered heterostructures can be estimated from the value of 
the normal diffusion constant, $\tau_{esc}~=~c^2/D_{\perp}$. For our sample 
we have the anisotropy factor $D_{\|}/D_{\perp}\cong 10^4$ and 
$D_{\|}=0.8 cm^2s^{-1}$ at $300\,K$. Then $\tau_{esc}\cong 4\cdot 10^{-11}s$ 
even at room temperature so that the condition $\tau_{esc} > \tau_{\varphi}$ 
may be really fulfiled at low temperatures.

\vspace{0.2in}

This research is supported by the Russian Program ``Topical Problems of 
Condensed Matter Physics'', Grant No.98004.


\newpage
\section{Sugnatures to the figures}

Fig. 1(a) The resistivity as a function of magnetic field at different 
temperatures for sample with $x = 0.18$.\\
(b) The surface conductivity as a function of ln$B$ for sample with $x=0.18$

Fig. 2 Logarithmic temperature dependence of the resistivity at $B > B_{c2}$ 
($x=0.18$).

Fig. 3 The surface conductivity as a function of magnetic field ($x = 0.12$).

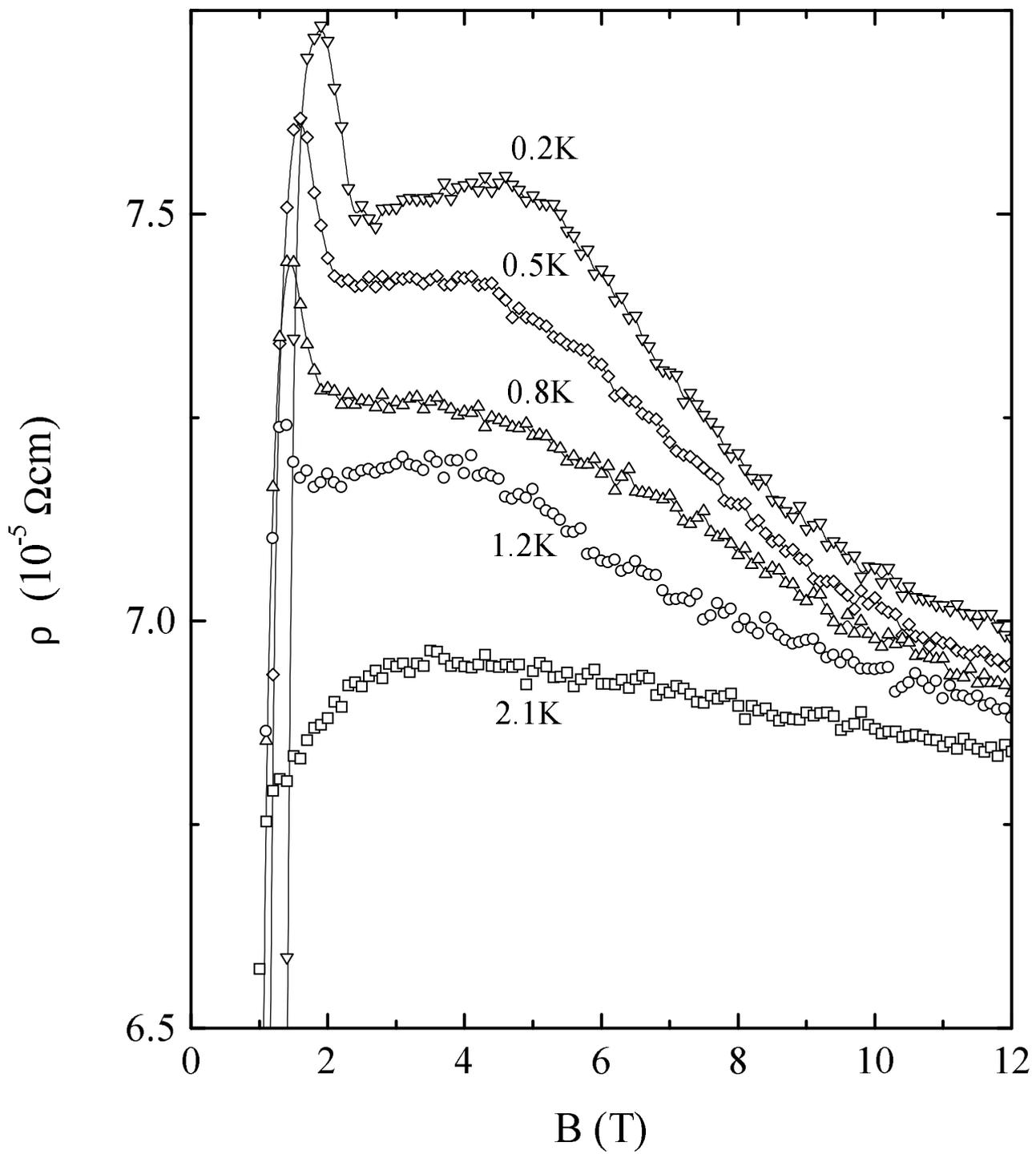
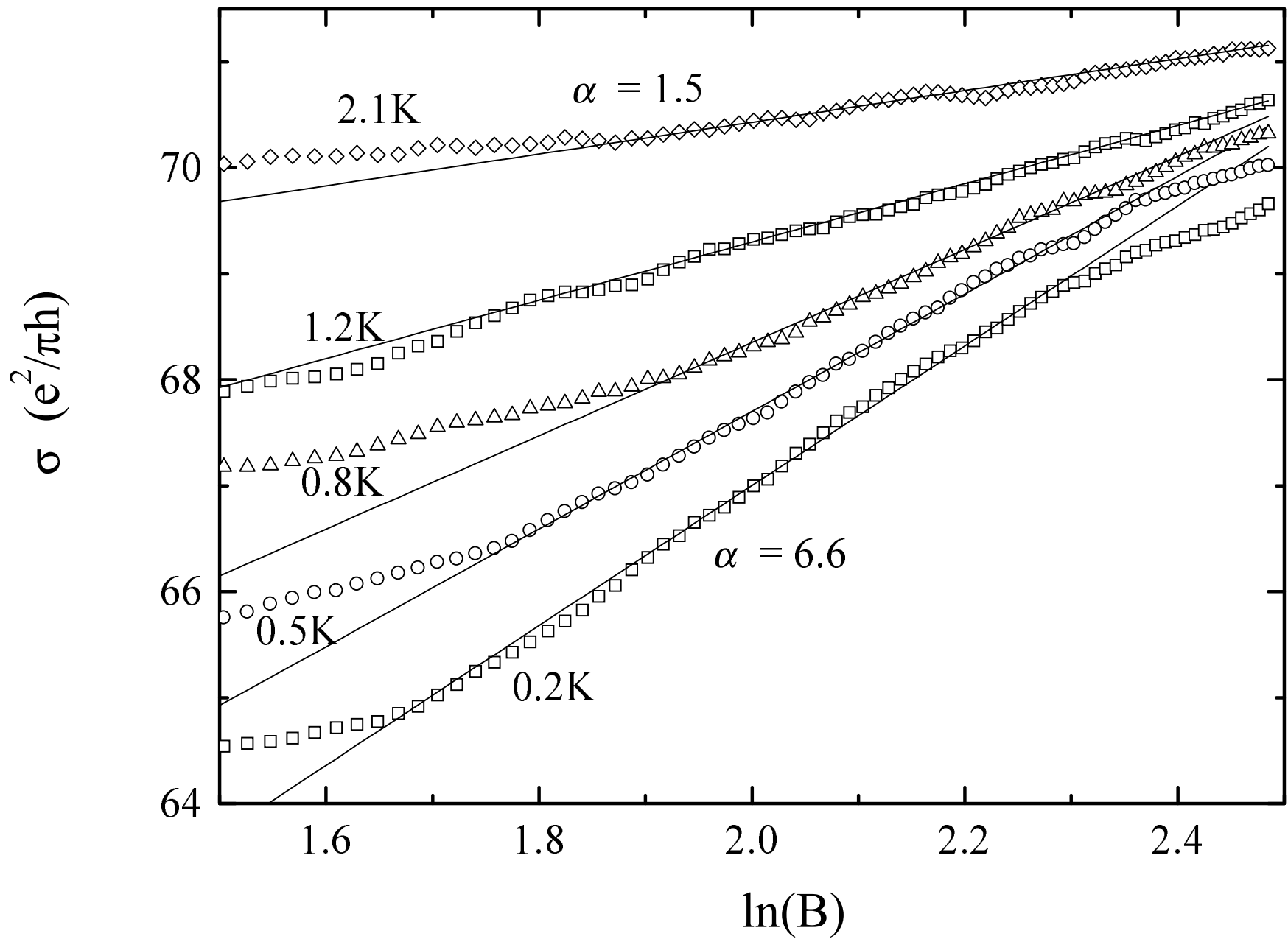
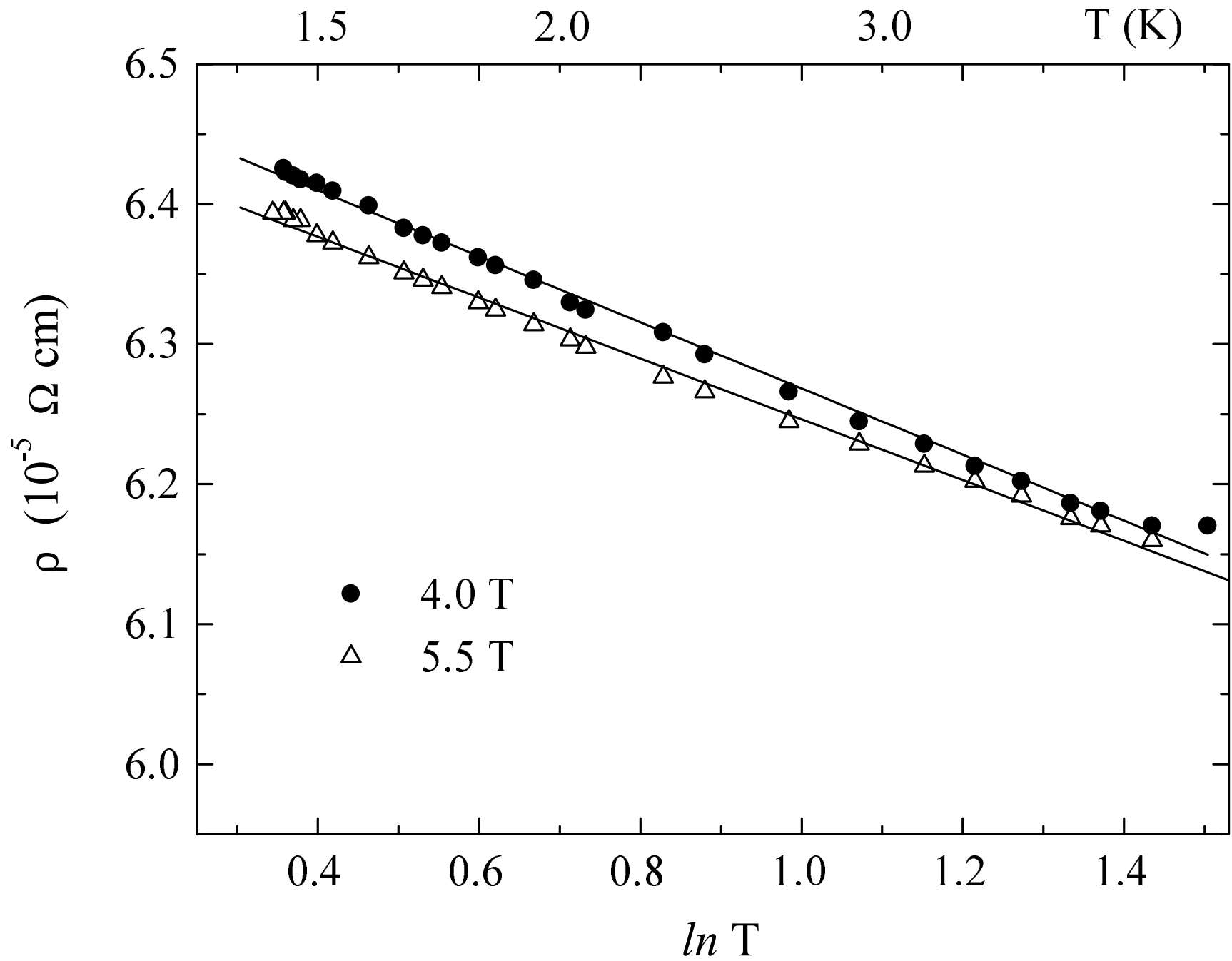
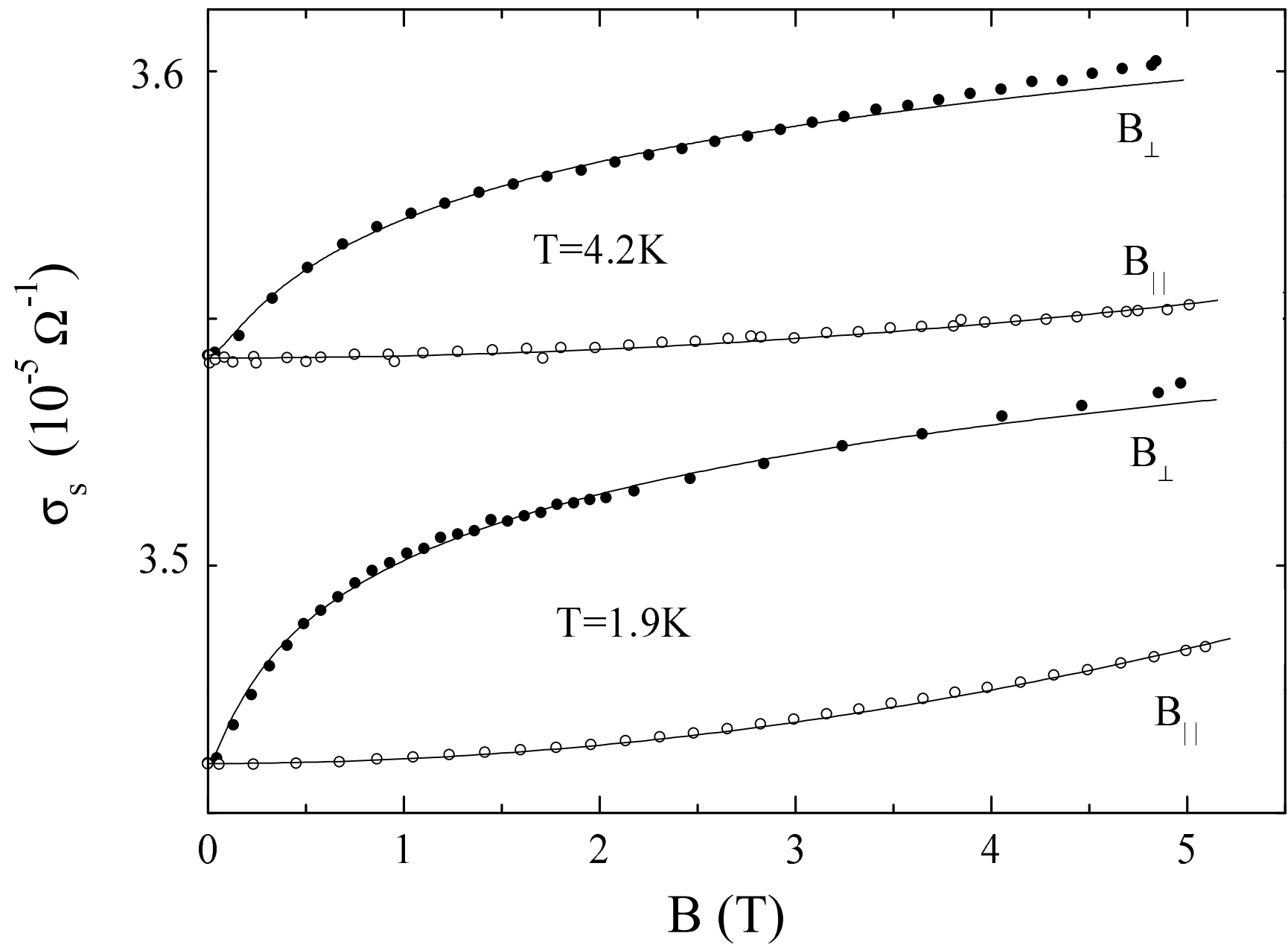
\end{document}